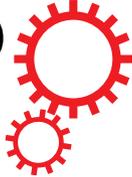

# Piezo Voltage Controlled Planar Hall Effect Devices

Bao Zhang[1], Kang-Kang Meng[2], Mei-Yin Yang[1], K. W. Edmonds[3], Hao Zhang[1], Kai-Ming Cai[1], Yu Sheng[1,4], Nan Zhang[1], Yang Ji[1], Jian-Hua Zhao[1], Hou-Zhi Zheng[1] & Kai-You Wang[1]

The electrical control of the magnetization switching in ferromagnets is highly desired for future spintronic applications. Here we report on hybrid piezoelectric (PZT)/ferromagnetic ($Co_2FeAl$) devices in which the planar Hall voltage in the ferromagnetic layer is tuned solely by piezo voltages. The change of planar Hall voltage is associated with magnetization switching through 90° in the plane under piezo voltages. Room temperature magnetic NOT and NOR gates are demonstrated based on the piezo voltage controlled $Co_2FeAl$ planar Hall effect devices without the external magnetic field. Our demonstration may lead to the realization of both information storage and processing using ferromagnetic materials.

Since the spin field effect transistor was proposed by Datta and Das in 1990s, due to the difficulties of spin injection and detection, until now there is still no efficient spin field effect transistor developed[1–3]. Recently, the spin Hall transistor based on two-dimensional electron gas has been demonstrated, in which the spins were generated optically rather than electrically in the semiconductor channel[4]. However, all-electrical manipulation of the magnetization is more desirable for spintronic applications. Approaches to control the spin of magnet bits electrically, such as spin-orbit torque[5–15], magneto-electrical coupling[16,17], voltage[18,19], polarized light[20] and piezo voltages[21,22] have been proposed based on ferromagnetic metals. Among them, the piezo voltage is one of the most effective methods to control the magnetization switching, in which a deformation of the crystal structure of the magnetic materials induces a change of the magnetocrystalline anisotropy which is directly related with the spin-orbit interaction in the crystal[23–28]. The control of the charge transport in semiconductors by piezo voltages has also been demonstrated for high speed piezotronics, which has been proposed for post Complementary Metal Oxide Semiconductor (CMOS) technology[29].

Here we demonstrate planar Hall effect devices in which the planar Hall resistances/voltages of ferromagnetic $Co_2FeAl$ devices can be tuned by piezo voltages from positive (negative) to negative (positive) effectively, which is associated with magnetization switching in the plane by 90°. Room temperature magnetic NOT and NOR gates are demonstrated based on the $Co_2FeAl$ piezo voltage controlled planar Hall effect devices without the external magnetic field. The simple device structure allows us to build large scale building blocks for future logics.

## Results

**The design of piezo voltage controlled planar Hall effect devices.** The Heusler alloy $Co_2FeAl$ films used in our planar Hall effect devices were grown by molecular beam epitaxy (MBE)[28,30]. The magnetization of the $Co_2FeAl$ is controlled by the uniaxial deformation induced by a piezo voltage applied to a PZT transducer, and the detection is provided by the planar Hall voltage across the Hall bar devices (See Methods and Supplementary S1). Schematic diagrams of the two $Co_2FeAl$ Hall bars with the respect to the GaAs crystal orientation along [100] and [010] axes are shown in Fig. 1a. We first fully magnetized the Hall bar devices along the [110] direction with magnetic field of 500 Oe (which is much larger than the coercive field), then swept the external magnetic field to zero. The measured $R_H$ with periodic piezo voltage ($U_P$) pulses applied to both devices is shown in Fig. 1b. A constant Hall voltage offset of the Hall device has been removed. Strikingly, the $R_H$ of the [100] orientation device periodically switched from $-0.12$ to $0.12\,\Omega$ with $U_P$ changing from 0 to $-30\,V$. In contrast, the $R_H$ of the [010] orientation device was periodically switched simultaneously from 0.12 to $-0.12\,\Omega$. The $R_H$ transitioned from negative low value to positive high value for sample [100], whereas the [010] orientated device changed from positive high to negative low under the same range of $U_P$, which is analogous to the n-FET and p-FET in CMOS technology. The advantage of this planar Hall effect device is that the $R_H$ (induced by planar Hall effect) changes sign during the tuning, whereas the conventional FET is switched on and off by accumulating and depleting the electrons (holes), while the resistance of the FET is always positive. In the piezo voltage control planar Hall effect devices, the changing of $R_H$ sign originates from the rotation of the magnetization vector respective to the electrical current under piezo voltages.

[1]SKLSM, Institute of Semiconductors, CAS, P. O. Box 912, Beijing 100083, People's Republic of China. [2]School of Materials Science and Engineering, University of Science and Technology Beijing, Beijing 100048, China. [3]School of Physics and Astronomy, University of Nottingham, Nottingham NG7 2RD, United Kingdom. [4]Department of Physics, School of Sciences, University of Science & Technology Beijing, Beijing 100048, China. Correspondence and requests for materials should be addressed to K.-Y.W. (email: kywang@semi.ac.cn)

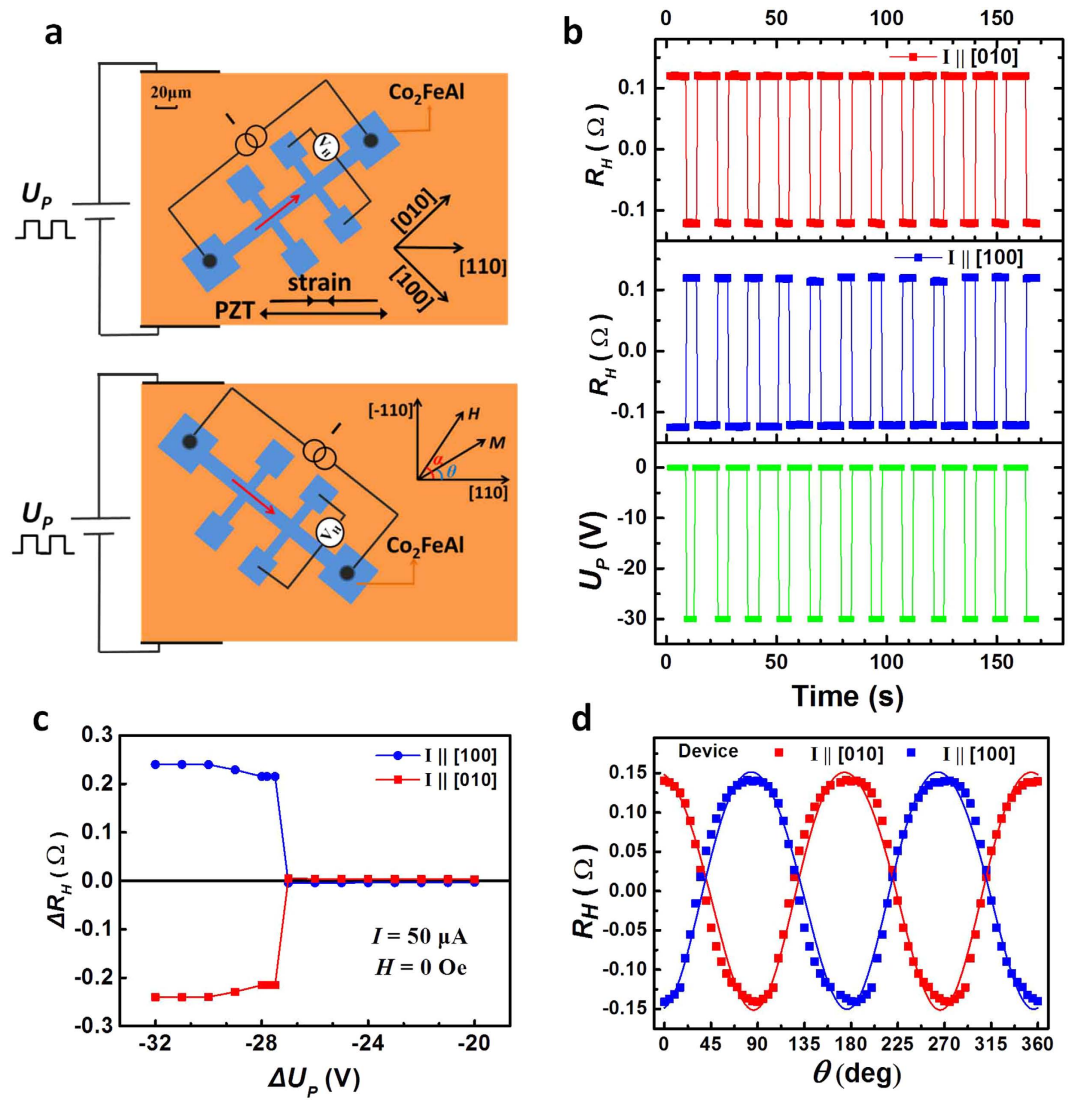

**Figure 1. The planar Hall resistance of Co$_2$FeAl device structure controlled by the piezo voltages.** (**a**) The schematic diagrams for measurements of planar Hall effect and application of piezo voltages for both the [010] and [100] orientated devices. (**b**) The periodic changes of the $R_H$ for both the [010] and [100] orientated devices with the periodic change of the piezo voltage between 0 V and −30 V without external magnetic field. (**c**) The change of the $R_H$ induced by a change of the piezo voltage from 0 V to certain values for both the [010] and [100] orientated devices. (**d**) The dependence of the $R_H$ for both the [010] and [100] orientated devices on the angle of magnetization under a fixed magnetic field at 2,000 Oe rotated in the plane, where the dots are the experimental results and the lines are the fitted results.

The change of the planar Hall resistance ($\Delta R_H$) on switching the piezo voltage from zero to certain values for both the [010] and [100] orientated devices is shown in Fig. 1c. The $\Delta R_H$ remains almost zero with switching the $U_P$ from 0 to values above −27 V, but a sharp jump and then a flat plateau were observed for both devices with switching the piezo voltages to a further negative value. Opposite values of $\Delta R_H$ (0.25 Ω and −0.25 Ω) were observed for devices along [100] and [010] orientations at the plateau. The dependence of the $R_H$ on the angle of magnetization (with respect to the [110] direction) is shown in Fig. 1d for both devices at zero $U_P$. The magnetization is rotated using an applied magnetic field of 2,000 Oe, which is large enough to force the magnetization vector to follow the field direction. The angular dependence of the $R_H$ can be fitted well using the single domain model[31], $R_H = \frac{1}{2}(R^\parallel_{sheet} - R^\perp_{sheet})\sin[2(\pm \pi/4) + \gamma]$, where $R^\parallel_{sheet}$ is the sheet resistance with the current parallel to the magnetization, $R^\perp_{sheet}$ is the sheet resistance with the current perpendicular to the magnetization, $\theta + \pi/4$ and $\theta - \pi/4$ represent the angle between the electrical current and the magnetization vector for [010] and [100] orientated devices respectively, and $\gamma$ is the deviation angle between the applied strain direction and [110] direction which is about 2°. The periodic $R_H$ for both devices has the same magnitude and frequency but with $\pi/2$ phase shift. The maximum magnitude of the Hall resistance occurs when $\theta = n\pi/2$, where n is an integer. The angular dependence of $R_H$ shows that switching magnetization by 90° induces a change $\Delta R_H$ of 0.27 Ω, which coincides with the large $\Delta R_H$ value in Fig. 1c. Thus, the piezo voltage can fully switch the magnetization of the Co$_2$FeAl devices by 90° in the plane.

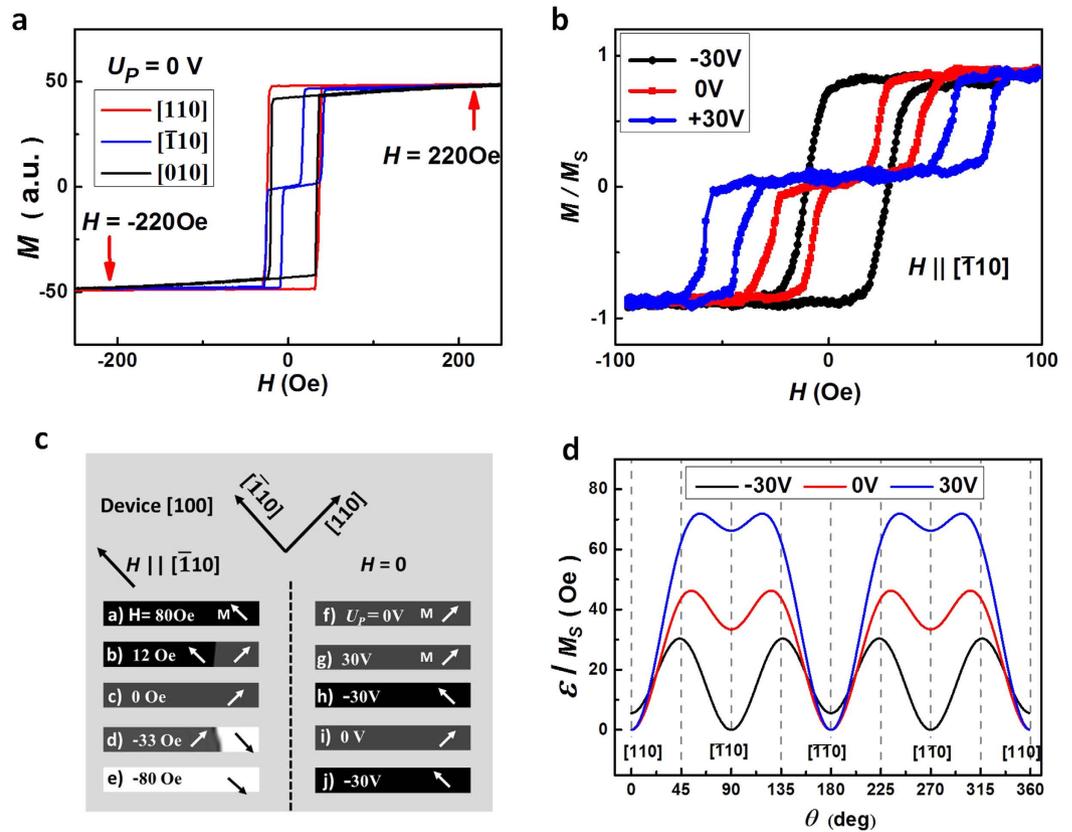

**Figure 2. The magnetic states of Co$_2$FeAl device controlled by the piezo voltages.** (**a**) The magnetic hysteresis loops of Co$_2$FeAl device measured by LMOKM with magnetic field applied in the [110], [$\bar{1}$10] and [010] directions with piezo voltage at zero. (**b**) The magnetic hysteresis loops with magnetic field in [$\bar{1}$10] orientation with piezo voltages at −30, 0 and 30 V. (**c**) The magnetic domain images (a–e) of the Co$_2$FeAl device without deformation during the magnetization reversal by applied magnetic field along [$\bar{1}$10] orientation. The magnetic domain images (f–j) of the Co$_2$FeAl device controlled by piezo voltages without external magnetic field. (**d**) The angular dependence of the magnetic energy density at zero magnetic field for the Co$_2$FeAl device with piezo voltages at −30, 0 and 30 V, where the minimum energy is in [110]/[$\bar{1}\bar{1}$0] direction for both 0 and 30 V (stretched) and it is switched to [$\bar{1}$10]/[1$\bar{1}$0] orientation with piezo voltage at −30 V (compressed).

**The magnetic properties controlled by the piezo voltages.** To obtain further insight into the switching behavior of the Co$_2$FeAl planar Hall effect devices, the magnetic hysteresis loops various $U_P$ were investigated using longitudinal magneto-optical Kerr microscopy (LMOKM). Figure 2a shows the Kerr rotation angle during the magnetization reversal at zero $U_P$ (without deformation) with magnetic field applied close to the in-plane major crystalline [110], [$\bar{1}$10] and [010] orientations, respectively. [110] orientation is the easy axis and the hard-axis-like behavior is seen for the [010] direction with saturation occurring around 220 Oe. However, the [$\bar{1}$10] orientation hysteresis loop shows a two-step switching with a continuous reversible rotation between the two steps. The observed magnetic hysteresis loops are the consequence of the superposition of a uniaxial and a fourfold anisotropy, where the uniaxial easy axis is along [110] orientation under zero strain ($U_P = 0$), and the fourfold easy axes are along [110] and [$\bar{1}$10] orientations[32,33]. To demonstrate the evolution of the magnetic anisotropy under piezo voltages, the hysteresis loops along [$\bar{1}$10] of Co$_2$FeAl with $U_P$ at zero and ±30 V are plotted in Fig. 2b. With $U_P = 30$ V, although the two-step jumping was also observed during the magnetization reversal, the field range of the two sharp jumps increased dramatically by more than a factor of two. The continuous reversible region between the two-step jumps increases with increasing the applied the piezo voltage (details in Supplementary Fig. S2), indicating the [$\bar{1}$10] axis becomes harder. With $U_P = -30$ V as shown in Fig. 2b, one step magnetization reversal along [$\bar{1}$10] orientation was observed, suggesting the magnetic easy axis is switched by 90° from [110] to [$\bar{1}$10] in the plane with $U_P$ at −30 V.

The magnetic domain images of the Co$_2$FeAl [100] orientated device without deformation were taken by LMOKM with magnetic field applied in [$\bar{1}$10] orientation, which are shown in Fig. 2c(a–e). The magnetization corresponding to [$\bar{1}$10], [110] and [1$\bar{1}$0] oriented domains are marked with arrows. Domain wall propagation could be detected during the sweeping of magnetic field around −12 Oe and 33 Oe, which was associated with the two-step jumps of the hysteresis loop along [$\bar{1}$10] in Fig. 2a. Figure 2c(f–j) represents the domain images of the device controlled by piezo voltages without magnetic field, which clearly shows the magnetization switched between [$\bar{1}$10] and [110] axes.

The 90° manipulation of magnetization in the piezo voltage controlled Co$_2$FeAl planar Hall effect device is due to an extra uniaxial anisotropy introduced by piezo voltages. The magnetic energy density of the system without deformation can be written as[34,35]:

$$\epsilon(\theta) = \frac{1}{4} K_C \sin^2(2\theta) + K_U \sin^2(\theta - \gamma) - H M_S \cos(\theta - \alpha) \tag{1}$$

where $\theta$ is the angle between magnetization and easy axis [110] direction, $\alpha$ is the angle between the external magnetic field and [110] direction, $K_C$ is cubic anisotropy, $K_U$ is the uniaxial anisotropy, $M_S$ is the saturated magnetization. The magnetic anisotropy constants can be obtained by analyzing the magnetic hysteresis loops along the uniaxial hard orientation, in the reversible region between the two-step jumps (details in Supplementary S3). The $K_C$ and $K_U$ were obtained to be $(108 \pm 5)M_S$ and $(41 \pm 2)M_S$, respectively. Then the angular dependence of the magnetic energy can be plotted, as shown in Fig. 2d. An additional strain-induced uniaxial magnetic anisotropy term $K_P \sin^2(\theta - \gamma)$ is added to the magnetic energy density equation (1) when $U_P \neq 0$, where $K_P$ has the same/opposite sign of $K_U$ at positive/negative piezo voltages. The obtained $K_P/M_S$ is $(26 \pm 1)$ and $(-46 \pm 4)$ Oe for $U_P$ at 30 and −30 V, respectively. The angular dependence of the magnetic energy at $U_P = \pm 30$ V is also plotted in Fig. 2d. The lowest energy state $\epsilon/M_S$ is along [110] orientation for $U_P$ at 0 and 30 V. However, the lowest energy is along [$\bar{1}$10] orientation at $U_P = -30$ V because the piezo voltage induced $K_P$ is larger than that of the $K_U$ but with opposite sign. The magnetic energy landscape in Fig. 2d suggests piezo voltages can switch the magnetic easy axis by 90°, which has also been confirmed using ferromagnetic resonance[28].

### The logic operation architectures of NOT and NOR logic gates.

In the present work, we propose and demonstrate NOT and NOR logic gates using the piezo voltage controlled planar Hall effect devices. Firstly, a high Hall voltage state (ON, digital '1') can be defined as an output voltage of $+5\,\mu$V or larger. A low Hall voltage state (OFF, digital '0') is defined as an output voltage of $+2\,\mu$V or smaller. For the piezo voltages, the 0 V and −30 V correspond to the logic state '0' and '1'. Based on the single piezo voltage controlled planar Hall effect device shown in Fig. 3a, the piezo voltages can effectively switch the magnetization between the [110] and [$\bar{1}$10] magnetic states, which function as the NOT gate and produce the planar Hall voltage output as shown in Fig. 1b. The truth table in Fig. 3b represents the NOT gate operation $Y = \overline{A}$. The NOR gate was built based on one [100] and one [010] planar Hall effect device connected as shown in Fig. 3c. The two piezo voltages ($U_{P1}$ and $U_{P2}$) control the [010] and [100] orientated devices separately. The magnetizations of two devices were preset to [110] orientation by external magnetic field. Then all the operations were executed without magnetic fields. Inputting the [0, 0] to the logic with both piezo voltages at zero sets the magnetization of both devices along [110], so that the $V_o$ is 11.8 $\mu$V and the output equal to 1. When the magnetizations of both devices is rotated to [$\bar{1}$10] direction by the piezo voltages, corresponding to the magnetic state [1, 1], the $V_o$ is $-11.6\,\mu$V $< 2\,\mu$V and output $= 0$. If only switching the magnetization of [100] or [010] device to [$\bar{1}$10] orientation, corresponding to the [1, 0] or [0, 1] magnetic states, the output is 0 since both $V_o$ are less than 2 $\mu$V (as shown in Fig. 3d). The non-zero value of Hall voltage at [1, 0] and [0, 1] states is because the two devices are slightly different. Figure 3e illustrates the NOR gate logic operations depending on the four separately measured values as defined in the Fig. 3d. The experimental *in-situ* measured output voltages were shown in the Supplementary Fig. S4, indicating that the logic devices have very good reproducibility.

### Discussion

In this experiment, we demonstrated that the piezo voltage is an effective way to control the magnetization and the resulting value of planar Hall voltage. When the piezo voltage $U_P > -27$ V, the magnetization is varied slightly and the Hall voltage change is small. However, when the piezo voltage $U_P < -27$ V, the Hall voltage has an obvious jump. It shows that the magnetization can be manipulated effectively by the piezo voltage.

The logic gates realized by the piezo voltage controlled planar Hall effect devices are desirable to be applied to the current computation technology. The time response of the piezo voltage controlled device has been investigated, where the rising and falling time are 220 $\mu$s and 70 $\mu$s, respectively (details in Supplementary S5). However, a smaller input voltage and a faster magnetization switching for the piezo voltage control should be achieved. Using the methodology of integrating the piezoelectric layers, the application of a smaller input voltage switching the magnetization is achievable by scaling down the devices to nanometer sizes[36]. Using the thin piezoelectric layer can also largely reduce the impedance thus largely increasing the device response speed. Our demonstration could pave a new way for the spin logic, realizing the information processing in only ferromagnetic metals.

### Methods

**Sample preparation.** The Heusler alloy Co$_2$FeAl (CFA) thin film was grown on semi-insulating GaAs (001) by using molecular beam epitaxial (MBE) technique at 280 °C. After deposition of 10 nm-thick CFA, the film was capped with an aluminum layer of 3 nm to avoid oxidation. The Hall bar devices along different in-plane major crystalline orientations were fabricated using standard photolithography and ion beam etching, where the device width is 20 $\mu$m and the distance between the neighbor arms is 80 $\mu$m. After polishing the GaAs substrate down to 100 μm thickness, the devices were bonded to the piezoelectric ceramic transducer (PZT) by glue. In order to ensure the deterministic switching of the $Co_2FeAl$ moment, all the devices were bonded ~2° from the [110] to [010] direction. The positive/negative voltage produces a uniaxial tensile/compressive strain in [110] orientation perpendicular to the stacks (the direction of strain is marked in Fig. 1a). The deformation of the $Co_2FeAl$ devices under the piezo voltage was found to be linearly changed with the applied piezo voltages, with tensile strain under positive piezo voltages and compressive strain under negative piezo voltages (See Supplementary Fig. S1).

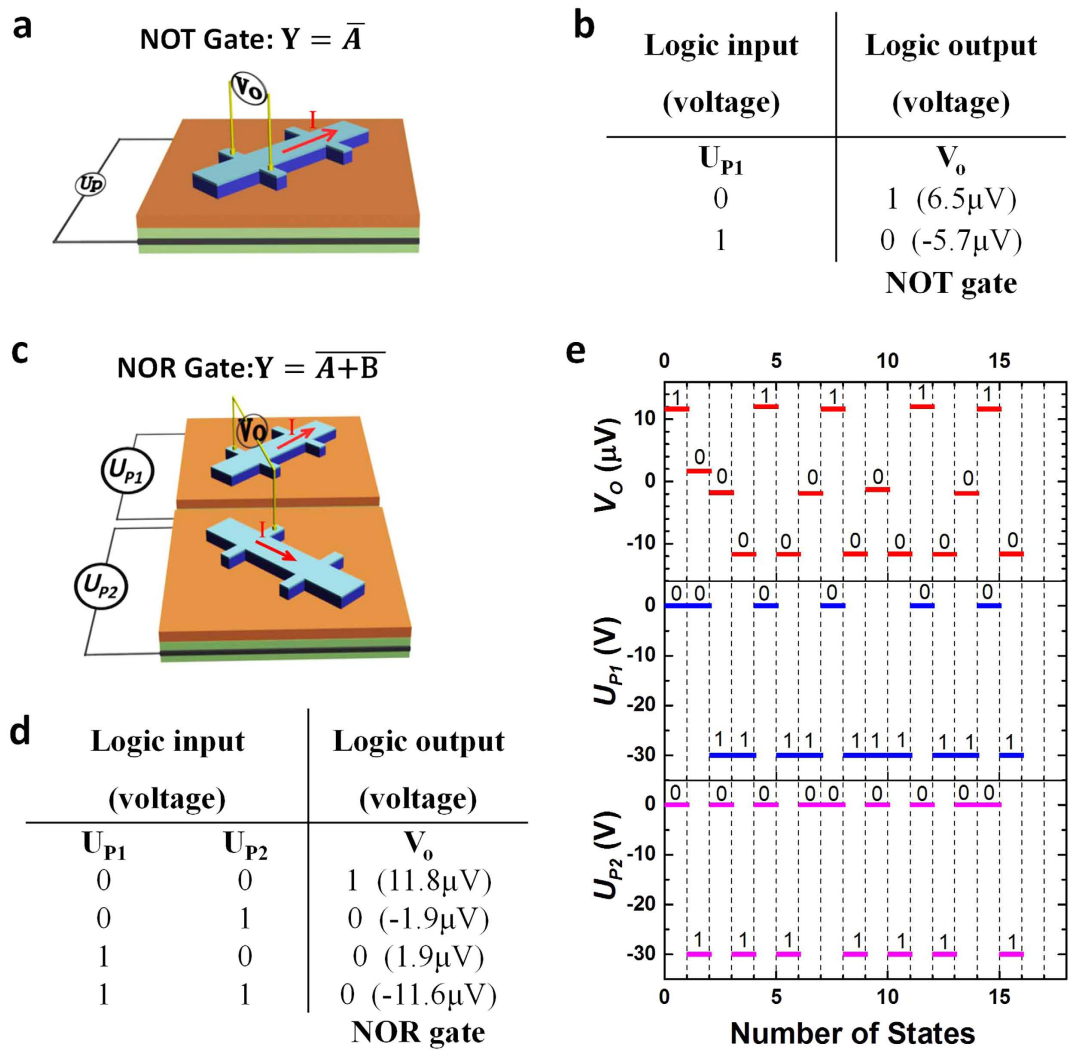

**Figure 3. Programmable logic operations demonstrated by a NOT and a NOR gate.** (**a**) The schematic diagram of a piezo voltage controlled [100] orientated $Co_2FeAl$ device built for NOT gate. (**b**) Truth table summary of the operation described in NOT gate. (**c**) The schematic diagram of piezo voltages controlled [010] and [100] $Co_2FeAl$ devices built for NOR gate, where the piezo voltages $U_{P1}$ and $U_{P2}$ are for the [010] and [100] devices, respectively. (**d**) Truth table summary of operation for the NOR gates with piezo voltage input. (**e**) Illustration of the NOR logic states operation with four separately measured output values as shown in the Fig. 3d.

**The measurements of the sample.** The magnetization vectors and the magnetic domains of the devices during magnetization reversal along in-plane orientations were measured by longitudinal magneto-optical Kerr microscopy (Nano MOKE3). At different deformation states controlled by the piezo voltages, a fixed current $I = 50\,\mu A$ passing through the channel was used to perform all the magnetotransport measurements, where the planar Hall voltages were detected using a Keithley nanovolt meter 2182. The time-dependent response of the device was measured by MOKE3 and the piezo voltages were applied by Agilent B1500A and the leading and trailing time are both 100 ns.

### References

1. Datta, S. & Das, B. Electronic analog of the electro-optic modulator. *Appl. Phys. Lett.* **56**, 665–667 (1990).
2. Koo, H. C. *et al.* Control of spin precession in a spin-injection field effect transistor. *Science* **325**, 1515–1518 (2009).
3. Sahoo, S. *et al.* Electric field control of spin transport. *Nature Phys.* **1**, 99–102 (2005).
4. Wunderlich, J. *et al.* Spin Hall effect transistor. *Science* **330**, 1801–1804 (2010).

5. Hayashi, M., Thomas, L., Moriya, R., Rettner, C. & Parkin, S. S. P. Current-controlled magnetic domain-wall nanowire shift register. *Science* **320,** 209–211 (2008).
6. Xu, P. *et al.* An all-metallic logic gate based on current-driven domain wall motion. *Nature Nanotech*. **3,** 97–100 (2008).
7. Liu, L. *et al.* Spin-torque switching with the giant spin Hall effect of tantalum. *Science* **336,** 555–558 (2012).
8. Bhowmik, D., You, L. & Salahuddin, S. Spin Hall effect clocking of nanomagnetic logic without a magnetic field. *Nature Nanotech*. **9,** 59–63 (2013).
9. Sinova, J. & Žutić, I. New moves of the spintronics tango. *Nat. Mater*. **11,** 368–371 (2012).
10. Locatelli, N., Cros, V. & Grollier, J. Spin-torque building blocks, *Nature Mater*. **13,** 11 (2014).
11. Wang, J., Meng, H. & Wang, J.-P. Programmable spintronics logic device based on a magnetic tunnel junction element. *J. Appl. Phys*. **97,** 10D509 (2005).
12. Miron, I. M. *et al.* Current-driven spin torque induced by the Rashba effect in a ferromagnetic metal layer. *Nature Mater*. **9,** 230–234 (2010).
13. Bijl, E. V. & Duine, R. A. Current-induced torques in textured Rashba ferromagnets. *Phys. Rev*. **B86,** 094406 (2012).
14. Liu, L., Lee, O. J., Gudmundsen, T. J., Ralph, D. C. & Buhrman, R. A. Current-Induced Switching of Perpendicularly Magnetized Magnetic Layers Using Spin Torque from the Spin Hall Effect. *Phys. Rev. Lett*. **109,** 096602 (2012).
15. Yang, M. *et al.* Spin-orbit torque in Pt/CoNiCo/Pt symmetric devices. *Sci. Rep*. **6,** 20778 (2016).
16. Leem, L. & Harris, J. S. Magnetic coupled spin-torque devices for nonvolatile logic applications. *J. Appl. Phys*. **105,** 07D102 (2009).
17. Cai, T. *et al.* Magnetoelectric coupling and electric control of magnetization in ferromagnet/ferroelectric/normal-metal superlattices. *Phys. Rev*. **B80,** 140415 (2009).
18. Maruyama, T. *et al.* Large voltage-induced magnetic anisotropy change in a few atomic layers of iron. *Nature Nanotech*. **4,** 158–161 (2009).
19. Zhu, J. *et al.* Voltage-Induced Ferromagnetic Resonance in Magnetic Tunnel Junctions. *Phys. Rev. Lett*. **108,** 197203 (2012).
20. Rubio-Marcos, F., Campo, A. D., Marchet, P. & Fernández, J. F. Ferroelectric domain wall motion induced by polarized light. *Nature Commun*. **6,** 6594 (2015).
21. Lei, N. *et al.* Strain-controlled magnetic domain wall propagation in hybrid piezoelectric/ferromagnetic structures. *Nature Commun*. **4,** 1378 (2013).
22. Lei, N. *et al.* Magnetization reversal assisted by the inverse piezoelectric effect in Co-Fe-B/ferroelectric multilayers. *Phys. Rev*. **B84,** 012404 (2011).
23. Li, Y., Luo, W., Zhu, L., Zhao, J. & Wang, K. Y. Voltage manipulation of the magnetization reversal in Fe/n-GaAs/piezoelectric heterostructure. *J. Magnetism and Magnetic Mater*. **375,** 148 (2015).
24. Ranieri, E. D. *et al.* Piezoelectric control of the mobility of a domain wall driven by adiabatic and non-adiabatic torques. *Nature Mater*. **12,** 808–814 (2013).
25. Rushforth, A. W. *et al.* Voltage control of magnetocrystalline anisotropy in Ferromagnetic-semiconductor/piezoelectric hybrid structures. *Phys. Rev*. **B78,** 085314 (2008).
26. Parkes, D. E. *et al.* Non-volatile voltage control of magnetization and magnetic domain walls in magnetostrictive epitaxial thin films. *Appl. Phys. Lett*. **101,** 072402 (2012).
27. Li, P. *et al.* Electric Field Manipulation of Magnetization Rotation and Tunneling Magneto resistance of Magnetic Tunnel Junctions at Room Temperature. *Adv. Mater*. **26,** 4320–4325 (2014).
28. Gueye, M. *et al.* Effective 90-degree magnetization rotation in $Co_2FeAl$ thin film/piezoelectric system probed by microstripline ferromagnetic resonance. *Appl. Phys. Lett*. **107,** 032908 (2015).
29. Newns, D., Elmegreen, B., Liu, X. H. & Martyna, G. A low-voltage high-speed electronic switch based on piezoelectric transduction. *J. Appl. Phys*. **111,** 084509 (2012).
30. Meng, K. K. *et al.* Magnetic properties of full-Heusler alloy $Co_2Fe_{1-x}Mn_xAl$ films grown by molecular-beam epitaxy. *App. Phys. Lett*. **97,** 232506 (2010).
31. Wang, K. Y. *et al.* Anisotropic magnetoresistance and magnetic anisotropy in high-quality (Ga,Mn)As films. *Phys. Rev*. **B72,** 085201 (2005).
32. Dumm, M. *et al.* Magnetism of ultrathin FeCo (001) films on GaAs (001). *J. Appl. Phys*. **87,** 5457–5459 (2000).
33. Qiao, S. *et al.* Magnetic and Gilbert damping properties of $L_{21}$-$Co_2FeAl$ film grown by molecular beam epitaxy. *Appl. Phys. Lett*. **103,** 152402 (2013).
34. Cowburn, R. P., Gray, S. J., Ferré, J., Bland, J. A. C. & Miltat, J. *et al.* Magnetic switching and inplane uniaxial anisotropy in ultrathin Ag/Fe/Ag(100) epitaxial films. *J. Appl. Phys*. **78,** 7210 (1995).
35. Wang, K. Y. *et al.* Spin reorientation transition in single-domain (Ga,Mn)As. *Phys. Rev. Lett*. **95,** 217204 (2005).
36. Solomon, P. M. *et al.* Pathway to the Piezoelectronic Transduction Logic Device. *Nano Lett*. **15,** 2391–2395 (2015).

### Acknowledgements

This work was supported by "973 Program" No. 2014CB643903 and NSFC Grant Nos 61225021, 11174272 and 11474272.

### Author Contributions

K.W. designed the whole experiments. B.Z., N.Z. and M.-Y.Y. fabricated the devices and performed the measurements. M.-Y.Y., K.-M.C., H.Z., K.W., Y.J., H.Z., Y.S. and B.Z. analyzed the data. K.-K.M. and J.-H.Z. provided the experimental materials. B.Z., K.W. and K.W.E. wrote this paper.

### Additional Information

**Supplementary information** accompanies this paper at http://www.nature.com/srep

**Competing financial interests:** The authors declare no competing financial interests.

**How to cite this article**: Zhang, B. *et al.* Piezo Voltage Controlled Planar Hall Effect Devices. *Sci. Rep.* **6**, 28458; doi: 10.1038/srep28458 (2016).